# Molecular dynamics simulation studies of 1,3-dimethyl imidazolium nitrate ionic liquid with water

*Oleg N. Starovoytov\**

Southern University and A&M College, Department of Computer Science, Baton Rouge, Louisiana 70813, United States

KEYWORDS ionic liquids, water, quantum mechanics, density functional theory, molecular dynamics, excess volume, density, and liquid structure

ABSTRACT The fundamental understanding of intermolecular interactions of ionic liquids with water represents a vital extent in predicting ionic liquid properties. Intermolecular or noncovalent interactions were studied for 1,3-dimethyl imidazolium [DMIM]$^+$ cation and nitrate [NO$_3$]$^-$ anion with water, employing quantum mechanics (QM) and molecular dynamics (MD) simulations. Extensive electronic structure calculations were performed first for molecular dimers, using various levels of theory and basis sets to pinpoint dimer optimized geometries and estimate binding energies in the gas phase. Many calculations resulted in planar dimer geometries for the cation-anion and anion-water pairs using 6-311++G(d,p) basis set. Dispersion corrected exchange correlation functionals resulted in more favorable binding energies for all tested pairs in comparison with energies obtained using Møller-Plesset second order perturbation theory (MP2). Molecular dynamics simulations were performed next using a revised multipolar polarizable force field (PFF). The effect of water on ionic liquids was evaluated in terms of thermodynamic properties. Thermodynamic properties included liquid densities $\rho$, excess molar volumes $\Delta V^E$, and liquid structures $g(r)$ as a function of water concentration. Densities of ionic liquid-water mixtures monotonically decrease while increasing the concentration of water. A negative excess volume is obtained for low water concentrations demonstrating strong intermolecular interactions of water with ionic liquid components. Liquid structures of ionic liquid - water mixtures revealed a tendency for anions to interact with cations at shorter intermolecular distances when water concentration is increased.

**Introduction**

Ethyl ammonium nitrate (EAN) salt was first described by Paul Von Walden in 1914.[1] It was mentioned as an organic ammonium salt (der organischen ammonium salze) with a low melting temperature. The organic salts with low melting temperatures, less than 373 K, can be referred to as room temperature ionic liquids (RTILs).[2] Ionic liquids are usually composed of organic heterocyclic cations and inorganic anions. The molecular structure of the cations can have a significant effect on ionic liquid melting and/or glass transition temperatures. Asymmetric cations lead to lower melting temperatures. Anions are recognized to be responsible for the solubility of ionic liquids in water.[2] Room temperature ionic liquids are an emerging class of materials that provide a number of essential thermodynamic properties for novel applications. Nevertheless, the thermodynamic properties of ionic liquids can be tuned by using a specific combination of cations and anions. Thermodynamic properties and potential applications of ionic liquids are thoroughly reviewed elsewhere.[3] Among many, imidazolium based ionic liquids are the most studied experimentally and computationally.[4]

The earliest 1-ethyl-3-methylimidazolium [EMIM]$^+$ based ionic liquids with tetrafluoroborate [BF$_4$]$^-$ and nitrate [NO$_3$]$^-$ anions were prepared and characterized in 1992.[5] Xray studies revealed that the nitrate anion forms a hydrogen bond with the hydrogen of C2 carbon of imidazolium ring (C2-H···O) in the solid state.[5] $^1$H NMR spectroscopy studies of imidazolium based ionic liquids in protic, ethyl ammonium nitrate, solvent also showed interaction of the nitrate anion with the hydrogen of C2 carbon atom of imidazolium ring.[6] Solvatochromic studies were performed for a number of ionic liquids to differentiate structural polarity correlations.[7] It was found that the ability of donating a hydrogen bond (HBD) strongly depends on the anion type in imidazolium based ionic liquids.

Quantum mechanical calculations though, using the G3MP2[8,9] level, revealed a pair structure for 1,3-dimethylimidazolium nitrate in the gas state.[10] The nitrate anion binds to the imidazolium ring in a planar configuration interacting with both methyl hydrogen atoms. Binding structures are similar, even with the presence of alkyl side-chain substituted groups.[10] Formation of hydrogen bonds was also obtained in DFT studies for the imidazolium based ionic liquids using various mono- and poly- anions.[11] Despite the formation of hydrogen bonds of the nitrate anion with the heterocyclic rings, these ionic liquids were characterized as hygroscopic.[5] The water molecule can also form a hydrogen bond with the nitrate and other anions, affecting the properties of ionic liquids. *Ab initio* calculations, classical molecular dynamics, and experimental studies are systematically carried out to better understand inter-molecular interactions and solvation of the nitrate anion in water.[12,13]

Molecular dynamics simulations are a complementary tool to experiments that can provide a fundamental understanding of inter-molecular interactions at the molecular level. A number of molecular dynamics simulations were performed to study the solvent effect on thermodynamics and structural properties of solutes.[14,15] Thermodynamic properties that are obtained from molecular dynamics simulations often rely on the quality of the force fields used in simulations.

Force fields are being systematically developed for the molecular dynamics simulations of ionic liquids to improve the accuracy of reproducible properties.[16,17,18] In this work, an AMOEBA[19,20] based multipolar polarizable force field is used that accurately describes electrostatic interactions at short intermolecular distances. The functional form of the potential is well described in a prior publication[16]. However, the multipolar polarizable force field was revised to improve the agreement of the torsional potential for the methyl hydrogen (C-N-C-H) with the quantum mechanical calculations at MP2/6-311G(d,p) level of theory, see the Supporting Information. All other force field parameters were taken without any corrections or adjustments including atomic polarizabilities. A series of molecular dynamics simulations were carried out using the revised force field for 1,3-dimethyl imidazolium [DMIM]$^+$ cation and nitrate [NO$_3$]$^-$ anion with water to determine thermodynamics properties. The chemical structures of 1,3-dimethyl imidazolium cation and nitrate anion are shown in Figure 1.

**Computational methods**

Molecular optimization was performed first for each dimer with the Hartree-Fock (HF) method [21,22] Møller-Plesset second-order perturbation theory[23], and a number of density functionals including functionals with van der Waals dispersion corrections following the proposed methodology[24]. Density functional calculations were carried out employing PBE[25], B3[26]LYP[27], M062X[28], B97XD[29], and including Grimme's dispersion correction[30,31,32] for planar and parallel dimer initial geometries. Calculations of binding energies are carried out next as a function of ion inter-molecular distances using theoretical levels mentioned above. Inter-molecular energies were corrected for the basis set superposition error[33,34] (BSSE) using the counterpoise correction[35] (CP) approach. The Gaussian G09 electronic structure software package[36] was used for quantum mechanics calculations.

Development of a polarizable force field greatly depends on the level of theory and a basis set used for the quantum mechanical calculations of molecular electronic structures and corresponding inter-molecular energies. The results of such calculations are usually referred to as a reference data set for the force field development. Therefore, the quantum mechanical calculations should be performed with an appropriate level of theory and a basis set to get accurate predictions of thermodynamic and other properties from molecular dynamics simulations. Recent studies showed that density functional theory empirically corrected for the London dispersion interactions outperforms the second order perturbation theory in the description of equilibrium geometries when compared to the geometries obtained using the coupled cluster with a full treatment of singles and doubles, and triple excitations complete-basis CCSD(T) level.[24] A number of quantum mechanics calculations were carried out to estimate the gas phase dimer geometries with respect to the dimer geometry obtained at MP2 level of theory. PBE, B3LYP, and M062X density functionals were involved in calculations as mentioned above. PBE is the general gradient approximation (GGA) functional for the exchange-correlation energy ($E^{xc}$), while B3LYP is the hybrid exchange functional ($E^{xc}_{hybrid}$) where HF exact exchange is mixed with GGA exchange and correlation.[26,27,37,38] The M062X functional was designed to take into account spin density, spin gradient, and spin kinetic energy density.[28] This functional is known as meta-GGA approximation.[39] Recent work showed that M062X gives better results in relative energies and molecular structures than the B3LYP functional for molecular systems involving dispersion and hydrogen bonding interactions.[39]

Studies of inter-atomic potentials of gases showed that density functionals could not adequately describe the London dispersion interactions.[40] Therefore, empirical corrections should be introduced to account for the London dispersion interactions. A few density functionals were developed that included empirical dispersion corrections to Kohn-Sham (KS) formalism[41]. These dispersion corrections are included as an additional term to the Kohn-Sham energy term as $E_{Total} = E_{KS-DFT} + E_{disp}$. The B97X-D functional is the dispersion-corrected density functional developed by M. Head-Gordon.[29] Approximation for the exchange correlation energy ($E^{xc}$) is carried out by accounting for the short- and long-range electron-electron interactions with the empirical corrections for the dispersion interactions as $E_{disp} = -\sum_{i=1}^{n-1}\sum_{j=i+1}^{n}\frac{C_6^{ij}}{R_{ij}^6}f_{damp}(R_{ij})$.

A damping function $f_{damp}(R_{ij})$ is introduced to provide the correct energy description at short and long inter-atomic distances.[29] Fast decay of a damping function at short inter-atomic distances allows for the dispersion interactions to be negligible within the intra-atomic distances.[30] Studies of inter-molecular interactions of water with aromatic compounds using a dispersion corrected B3LYP functional showed an improved description of inter-molecular energies using an atomic correction scheme.[42] Here, the B3LYP density functional with Grimme's D3 dispersion correction[31] (B3LYPD3) is used in calculations for comparison purposes. All calculations are performed with various basis sets. Pople[43] and Dunning[44] types of basis sets are used in calculations to study the effect of the basis set size on the inter-molecular structures.

**Geometry optimization**

Dimer geometry optimization is carried out first using various levels of theory and basis sets. Initial structures are set up as "planar" and "parallel" for the cation/anion and anion/water dimers, see the Supporting Information. Using PBE, M062X, and B3LYPD3 functionals resulted in various dimer geometries depending upon a basis set used in calculations for the initial planar cation/anion pair. The nitrate molecule makes hydrogen bonds with C2 hydrogen and a methyl hydrogen. Planar gas phase geometry is obtained for the dimer using HF, MP2, B3LYP, and B97XD levels of theory with 6-311G(d,p) basis set. Addition of the diffuse functions resulted in planar geometries for all levels but M062X. Interactions of the cation and anion are shifted to the longer inter-molecular distances, see Table 1. The inter-molecular energies are more prominent when polarization functions are used on heavy atoms and hydrogens. Calculation of inter-molecular energies with the diffuse functions yields smaller variations. Inter-molecular distances are close when calculations are performed using the MP2 level and B97XD density functional. However, inter-molecular interaction energy is ~3.2 kcal/mol more favorable when B97XD density functional is used. Using a correlation consistent cc-pVDZ basis set resulted in out of plane dimer geometries using the HF, MP2, PBE, B3LYP, and B3LYPD3 levels. Planar geometries are obtained while using the cc-pVTZ basis set. Optimizing parallel cation/anion initial geometries resulted in all parallel geometries with some variations in inter-molecular distances. One of the oxygen atoms of the nitrate anion tends to stay closer to the C2 hydrogen, making a hydrogen bond.

Geometry optimization for the nitrate anion/water dimer was carried out in a similar manner. Planar geometries were obtained using all levels of theory and basis sets except at the HF level. The water molecule tends to interact with two oxygen atoms of the nitrate anion, making two hydrogen bonds. In some dimer geometries, one hydrogen atom interacts with one oxygen atom of the nitrate anion when the HF level of theory is applied. Inter-molecular distances are consistently longer while calculating with 6-311++G(d,p) basis set, see Table 1. Corresponding inter-molecular energies are systematically less attractive in comparison with the 6-311G(d,p) basis set. Inter-molecular distances do not change when Dunning basis sets are used with the triple

(T) and quadruple (Q) valence functions augmented with the diffuse functions and calculated with the B97XD density functional, see the Supporting Information.

**Inter-molecular interactions**

Inter-molecular energy calculations are carried out next in a similar manner as previous calculations. Optimized geometry at MP2 level of theory and the 6-311G(d,p) basis set was used to study interaction energies as a function of inter-molecular separation distance. This level of theory was chosen to comply with the AMOEBA force field development methodology. Optimized molecular geometries at the MP2 level with 6-311G(d,p) basis set are illustrated in Figure 2. These geometries are used as initial structures for calculating binding energies. The nitrate $[NO_3]^-$ anion interacts with the $[DMIM]^+$ cation along the N···H-C2 vector as shown in Figure 2(a). The nitrate anion lies within the plane of the heterocyclic ring of the cation keeping its trigonal planar geometry. Oxygen atoms of the nitrate anion make hydrogen bonds with the methyl hydrogen atoms of 1,3-dimethyl imidazolium cation. Methyl hydrogen atoms lie within the plane of the imidazolium ring that corresponds to the lowest energy conformer in an isolated state. A similar optimized structure was obtained using G3MP2 theory.[10]

The water molecule interacts with one of the nitrogen atoms of the cation ring along the O···N vector as schematically shown in Figure 2(b). The water molecule is slightly turned toward the proton of the C2 carbon of the imidazolium ring. It looks like water makes a hydrogen bond with that hydrogen atom of the imidazolium ring. It is not positioned within the same plane of imidazolium ring as in the case of $[DMIM]^+$ - $[NO_3]^-$ interactions. In fact, the H-O-H plane is somewhat perpendicular to the plane of the imidazolium ring.

The water molecule interacts with the nitrate anion along the O-N⋯O vector as shown in Figure 2(c). Like [DMIM]$^+$ - [NO$_3$]$^-$ interactions, the water molecule lies within the same plane of the nitrate anion indicating "planar" geometry. Hydrogen bonds are made with the anions' oxygen atoms and hydrogen atoms of the water molecule.

Total intermolecular potentials are plotted in Figure 3(a-c) for all dimers. Red open circles correspond to the inter-molecular energies obtained from calculations using quantum mechanics. Black lines correspond to the energies reproduced employing a revised multipolar polarizable force field. Inter-molecular distances are considered near the molecular equilibrium separation, full potentials can be found in the Supporting Information. Multipoles for the [DMIM]$^+$ cation and [NO$_3$]$^-$ anion were taken from the previously published multipolar PFF force field.[16] Gaussian distributed multipole analysis (GDMA[45,46,47]) was used for the fitting of electrostatic interactions and marked as GDMA G1. Intra-molecular parameters for the cation and anion are also taken from that reference. Vibrational frequency analysis was completed for the nitrate anion resulting in underestimation of frequencies in comparison to the experiment. Polarization energies were well described with the GDMA G1 set of parameters (atomic polarizabilities) at short and long separation distances, but underestimated energies at the equilibrium intermolecular distance for the "parallel" geometry.[16] The force field parameters for the [NO$_3$]$^-$ anion were taken as described in the previous section. The AMOEBA[19,20] water model was taken without any adjustments. Previous parametrization of the multipolar polarizable force field was completed using "parallel" dimer geometry when the nitrate anion was located on top of the imidazolium ring.[16]

**Inter-molecular interactions of [DMIM]$^+$ cation with [NO$_3$]$^-$ anion**

Figure 3(a) shows total inter-molecular potential obtained for the [DMIM]$^+$ - [NO$_3$]$^-$ interactions. The minimum interaction energy is 92.25 kcal/mol at the C⋯N equilibrium distance of 3.35 Å. The revised multipolar polarizable potential underestimates these interactions by ~1.1 kcal/mol at the minimum, if compared to the MP2 level of theory. The B3LYP density functional performs similarly to the MP2 level of theory, getting slightly more favorable interactions. The B97XD density functional provides ~3.0 kcal more attractive interactions than MP2 and B3LYP. The B3LYP with the Grimme's D3 correction estimates the binding energy at -96.02 kcal/mol.

**Inter-molecular interactions of [DMIM]$^+$ cation with H$_2$O**

Interaction energies are shown for the [DMIM]$^+$ with water in Figure 3(b). The interaction energies for the cation/water pair are significantly lower if compared to the cation/anion pair. The multipolar polarizable force field tends to underestimate inter-molecular binding energies of water with the cation by ~1.0 kcal/mol. The B3LYP density functional gives binding energies that are comparable with the MP2 second order perturbation. Binding energies are more favorable when B97XD and B3LYPD3 density functionals are used, a small energy difference can be seen between the two.

**Inter-molecular interactions of [NO$_3$]$^-$ anion with H$_2$O**

Interaction energies are shown for the [NO$_3$]$^-$ anion - water pair in Figure 3(c). These energies are more favorable than for the cation/water pair. The multipolar polarizable force field overestimates these interactions by ~3 kcal/mol. Binding energies estimated using various levels of theory follow the same order as previous calculations. The minimum interaction distance is shifted to a shorter distance of 3.1 Å. One of the possible reasons for the energy overestimation could be the atomic polarizabilities of the nitrate anion.

Molecular/atomic polarizabilities play an important role in inter-molecular interactions such as ion solvation. High quality of isotropic atomic polarizabilities were obtained from experiment based on the Lorenz-Lorenz relation.[48] Isotropic atomic polarizabilities were parametrized based on four different polarizable models and experimental data.[49] Computational studies were carried out to parametrize the nitrate atomic polarizabilities in response to the electric field of the lithium cation as a function of lithium-ion separation distance employing the Kitaura-Morokuma[50,51] decomposition approach at the HF level of theory.[15] A good description of polarization energies was attained within the whole range of inter-molecular distances, up to 5.5 Å. However, the atomic polarizabilities for the nitrogen and oxygen atoms of the nitrate anion are different from those obtained for the ionic liquids.[16] The results of parameterizations are summarized in Table 2.

*Ab initio* and experimental studies were performed to determine the molecular polarizabilities for the nitrate[12,52] and other anions[53]. Atomic polarizabilities for the nitrate anion were determined based on inter-molecular interactions of $[NO_3]^-$ $(H_2O)_3$ clusters in non-planar $C_3$ symmetry isomer calculated at B3LYP and MP2 levels of theory and aug-cc-pVDZ basis set.[12] The values for isotropic polarizabilities of 4.67 - 4.99 Å$^3$ were reported depending upon the theoretical level and basis set. Contribution of the central nitrogen atom to the average isotropic molecular polarizability was calculated to be very small, even negative. Molecular polarizability of 4.48 Å$^3$ (30.2 a.u.$^3$) was reported for the nitrate anion in aqueous solutions.[52] The atomic polarizabilities were also reported for the nitrogen (1.698 Å$^3$) and oxygen (1.144 Å$^3$) atoms from *ab initio* analysis of ionic liquids.[53] These values will give an average molecular polarizability of 5.13 Å$^3$, which is very close to that determined from isomers. Assuming additivity of the atomic polarizabilities, a total molecular polarizability of 3.584 Å$^3$ is obtained for the multipolar PFF.[16]

Despite that the polarizable model of the multipolar force field underestimates the molecular polarizabilities reported earlier from *ab initio* and experimental studies, the anion/water binding energy is significantly overestimated. Four sets of atomic polarizabilities are then used to perform molecular mechanical calculations and to compare intermolecular energies. These calculations showed only ~0.3 kcal/mol energy gain when going from a molecular polarizability of 0.00 Å$^3$ to 5.13 Å$^3$. The result indicates that electrostatic and non-bonded interactions should be considered in detail to improve the agreement of the PFF with QM data.

**Molecular dynamics simulations**

Molecular dynamics (MD) simulations were performed for ionic liquids using the tinker 4.3 simulation package[54]. The AMOEBA[19,20] force field was used to carry out MD simulations using a cubic simulation cell with applied periodic boundary conditions. The standard shake algorithm was employed to constrain the bond lengths. The ionic liquid systems were set up in a periodic simple cubic (SC) lattice that included 216 ionic pairs (3672-4536 atoms depending upon the system). Energy minimization was performed using a conjugate gradient minimization algorithm to reduce energetic strains. Simulations were carried out first in the NVT ensemble to heat up the system to 408 K. Isobaric-isothermal NPT simulations were performed next until steady state conditions are reached with the integration time step of 1 fs. Equilibration period was carried for more than 1 nanosecond (ns) period. Production trajectories were generated for more than 5 ns depending upon the system size. The cutoff radius was 9.0 Å for non-bonded interactions. Long range electrostatic interactions were computed using Ewald summation[55] with an 8 Å direct cutoff. The Beeman integration algorithm[56] was employed. The Berendsen thermostat and barostat[57] were used to control temperature and pressure with a relaxation time of 1.0 ps.

**Results and discussion**

Thermodynamic properties of pure water and the 1,3-dimethyl imidazolium nitrate are evaluated first from molecular dynamics simulations using a multipolar polarizable force field. Liquid densities of the pure water and ionic liquids are plotted as a function of temperature in Figure 4. Water liquid densities are described within 1.0% error in comparison with experimental data, see Figure 4 (a). The AMOEBA[19,20] water model well described experimental liquid densities at all temperatures except the density at 408 K. However, molecular dynamics simulations at 408 K were not corrected for the water saturation conditions. A multipolar polarizable force field gives linear dependence of the densities as a function of temperature, see Figure 4 (b). The density of the 1,3-dimethylimidazolium nitrate is underestimated by ~7.0 % in comparison to the experimental data. Experimental density of 1.36 g/cm$^3$ had been obtained for the crystal of 1,3-dimethylimidazolium nitrate.[58] The melting temperature of the 1,3-dimethylimidazolium nitrate was determined to be 344 K as indicated in Figure 4 (b). Experimental data for the liquid densities of the 1-ethyl 3-methyl imidazolium nitrate and 1-methylimidazolium nitrate are also included in the plot for comparison.[10] Liquid densities of the 1,3-dimethylimidazolium nitrate are in proximity to experimental densities of the 1-ethyl 3-methyl imidazolium nitrate. The experimental point at 294 K corresponds to the 1-methyl imidazolium nitrate with a value of 1.25 g/cm$^3$.[10] Linear dependence of densities is obtained as a function of temperature. The molecular volumes of the 1,3-dimethylimidazolium nitrate pair were estimated as a function of temperature. The results of these calculations are summarized in Table 3. The molecular volume of the 1,3-dimethylimidazolium nitrate is estimated to be only ~1.5 Å$^3$ larger at a temperature of 363 K. The predicted average volume at 408 K is ~7.51 Å$^3$ larger than volume at 353 K, indicating the presence of strong electrostatic interactions. Average molecular volumes were not calculated for the temperatures below the melting temperature, as these volumes would correspond to the supercooled regime. A series of molecular dynamics simulations were performed for ionic liquids-water mixtures at various water concentrations. All ionic liquids/water simulations were performed at 353 K to stay above the melting temperature of the 1,3-dimethylimidazolium nitrate, and, at the same time, to stay in the liquid regime for the pure water. The liquid density of water at 353 K is in good agreement with experimental data as shown in Figure 4 (a). Melting temperature ($T_m$) for the 1-ethyl-3-methyl imidazolium nitrate is ~311 K and 1-methylimidazolium nitrate is ~339 K.

**Liquid densities and excess volume for ionic liquid - water solutions**

Here, the intermolecular interactions are studied by calculating the liquid densities and excess molar volume as a function of water concentration. The excess molar volume $\Delta V^E$ of the 1,3-dimethylimidazolium nitrate - water solutions was determined as indicated by the equation below.

$$\Delta V^E = V_{mixture} - (x_{solute}V_{solute} + x_{solvent}V_{solvent})$$

Where $V_{mixture} = (x_{solute}V_{solute} + x_{solvent}V_{solvent})/\rho_{mixture}$ is the molar volume of the mixture. $V_i = (M_i x_i)/\rho_i$ is the molarity of the pure components (ionic liquid and water), $M_i$ is the molecular weights of the pure components, and $x_i$ is the mole fraction of the ionic liquid or water in the mixture.

Liquid density and excess molar volumes of ionic liquid - water binary mixtures are given in Figure 4. The liquid density is monotonically decreasing as water is being added to the mixture, see Figure 4 (a). Previous studies indicated that magnitude and sign of the excess volume directly depends on the type of anion in a system.[59] Competing effects can be a result of the excess molar volume. Positive excess volume indicates a weakening of the self-association of water or hydrogen bonds or ion coulomb attraction or van der Waals association between the components of ILs. A negative effect can be due to the association of ion-polar group attraction. This will bring different species closer together, smaller molecules fitting in between larger molecules. These competing effects can be understood better by looking at the partial molar volumes. Any association between the IL and water must compete with the two relatively strong self-associations of ion-ion and the H-bonds of water. Studies of excess volume for the 1-ethyl-3-methylimidazolium acetate with water were carried out at different temperatures. Complete miscibility was reported for this ionic liquid with water.[60] Negative excess volume was reported at all water concentrations, having an excess minimum of -1.52 kcal/mol at ~0.6.

**Liquid structures for the ionic liquid - water mixtures**

In order to evaluate inter-molecular structures, the radial distributions are built. Center-of-mass radial distribution functions (RDFs) were calculated for the pure 1,3-dimethylimidazolium nitrate and 1,3-dimethylimidazolium nitrate + water mixtures at 353 K as shown in Figure 6. Radial distributions for the 1,3-dimethylimidazolium nitrate liquid are shown in Figure 6 (a) and for 1,3-dimethylimidazolium nitrate + water mixtures in Figures 6 (b-d). Three weight fractions of water were selected for comparison. Radial distributions for the anion/water and cation/water are given in Figures 6 (c and d) respectively.

The first peak of the radial distribution function for the 1,3-dimethylimidazolium nitrate pair can be seen at a distance of 5 Å, see Figure 6 (a). Four nitrate anions can be found in the first coordination shell within a distance of 7 Å. One peak distribution is obtained for the $[NO_3]^-$ - $[NO_3]^-$ pair, at a distance of 6.8 Å. Two peak distributions can be seen for the $[DMIM]^+$ - $[DMIM]^+$ pair, indicating a prominent intermolecular structure. The location of the first peak can be found at a distance of 3.9 Å and the second at 7.5 Å. The structure that corresponds to the first peak could be described as a "parallel stacked", when two imidazolium rings can be found in parallel orientation close to each other. The radial distribution for 1,3-dimethylimidazolium nitrate pair is somewhat similar to the pure IL at low water fraction (0.10), see Figure 6 (b). The first peak can be seen at a distance of 5 Å. Increasing water content makes the distributions broader but the position of the distribution peak is still unchanged. The probability of finding an anion around a cation is decreasing, as indicated by obtaining lower distribution peaks.

These results suggest that the cation interacts with an anion at a preferable distance despite the amount of water solvent present in a system. Coordination numbers are monotonically decreasing with an increasing water fraction, reaching a value of less than 1 for a water fraction of 0.88. There is one narrow peak for $[NO_3]^-$ - $H_2O$ distributions at a distance of 3.46 Å, see Figure 6 (c). This distance is well associated with the equilibrium interaction distance in the gas phase. The narrow distribution implies an ordered intermolecular structure. The water coordination number increases with the water content, reaching ~10 water molecules around a nitrate anion in the first coordination shell when the water fraction equals to 0.88. The liquid structure of $[DMIM]^+$ - water is presented in Figure 6 (d). The location of the first peak can be found at an intermolecular distance of 4.48 Å. It can be seen that water molecules interact with the imidazolium ring at distances closer than anions. The first distribution is broad with a small shoulder getting flat as the water content is increasing.

**Conclusions**

Inter-molecular interactions were studied for 1,3-dimethylimidazolium cation and nitrate anion with water. The water molecule interacts with the imidazolium ring through the nitrogen-oxygen intermolecular association while the nitrate anion makes two hydrogen bonds with water in the gas phase. The MP2 level of theory gives inter-molecular equilibrium distances close to those obtained with B97XD when 6-311++G(d,p) basis set is used. A multipolar polarizable force field tends to underestimate binding energies at equilibrium inter-molecular distances for the cation/anion and cation/water pairs. Inter-molecular binding energies are greatly overestimated for the anion/water pair. Density functionals with the London dispersion correction give the most favorable binding energies over the hybrid and meta-GGA functionals. The order of binding energies can be set as HF < MP2 < B3LYP < B97XD < B3LYPD3 for all studied pairs.

Molecular dynamics simulations have been performed for [DMIM][NO$_3$] - water mixtures at 353K using a revised multipolar polarizable force field. Liquid densities of these mixtures are in good agreement with the experimental densities of [EMIM][NO$_3$] - water mixtures. Negative excess volume has been obtained for the low water concentrations and positive at high concentrations. Radial distribution functions showed a tendency of the nitrate anion to interact closer to the cation when water concentration is increased. Inter-molecular interactions of cations and anions are reduced at high water concentrations due to the ion solvation.

**FIGURES**

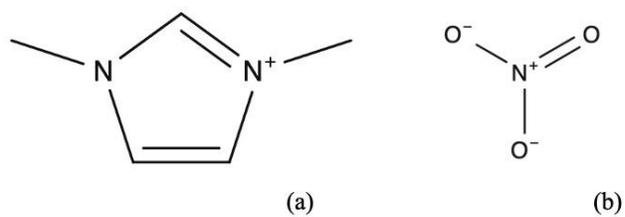

(a)          (b)

**Figure 1.** Chemical structures of 1,3-dimethyl imidazolium cation [DMIM]$^+$ (a) and nitrate anion [NO$_3$]$^-$ (b). Hydrogen atoms are not shown for clarity. Chemical structures were visualized using the MolView online resource.

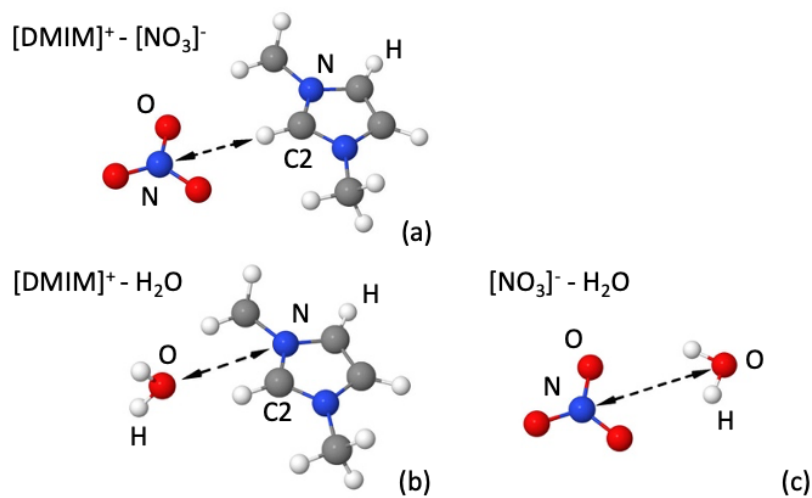

**Figure 2.** Schematic representations of the testing configurations are shown for the calculation of binding energies for the 1,3-dimethylimidazolium cation - nitrate anion pair (a), 1,3-dimethylimidazolium-water pair (b), and for the nitrate-water pair in $C_{2v}$ symmetry (c). The geometries are optimized at the MP2/6-311G(d,p) level of theory in the gas phase. The black dashed arrows represent testing directions.

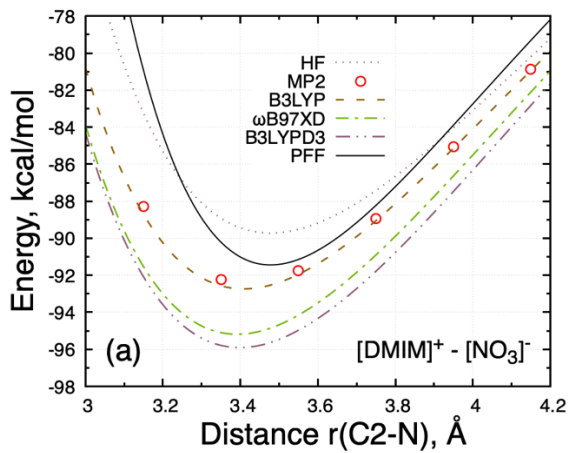

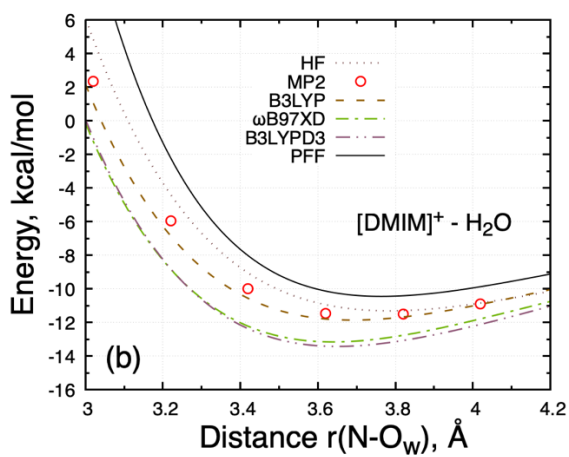

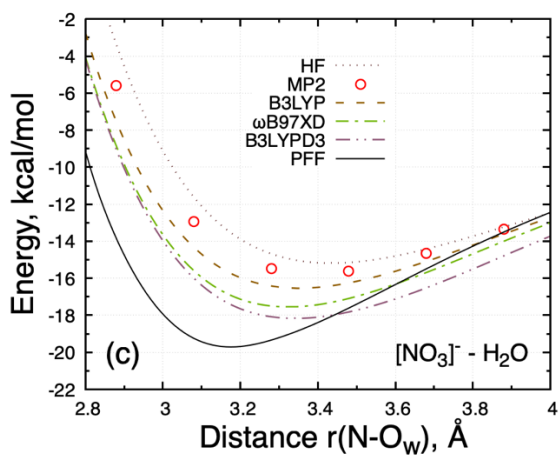

**Figure 3.** Total intermolecular potentials are shown as obtained from QM calculations (red circles) and revised multipolar PFF potential (black lines) as a function C-N⁻, N-O$_w$, and N-O$_w$ separation distances in the gas phase. Schematic representations of the testing configurations are shown for the calculation of binding energies of the 1,3-dimethylimidazolium cation with the nitrate anion (a), 1,3-dimethylimidazolium with water (b), and the nitrate anion with water (c). The geometries are optimized at the MP2/6-311G(d,p) level of theory in the gas phase. The black dashed arrows represent testing directions.

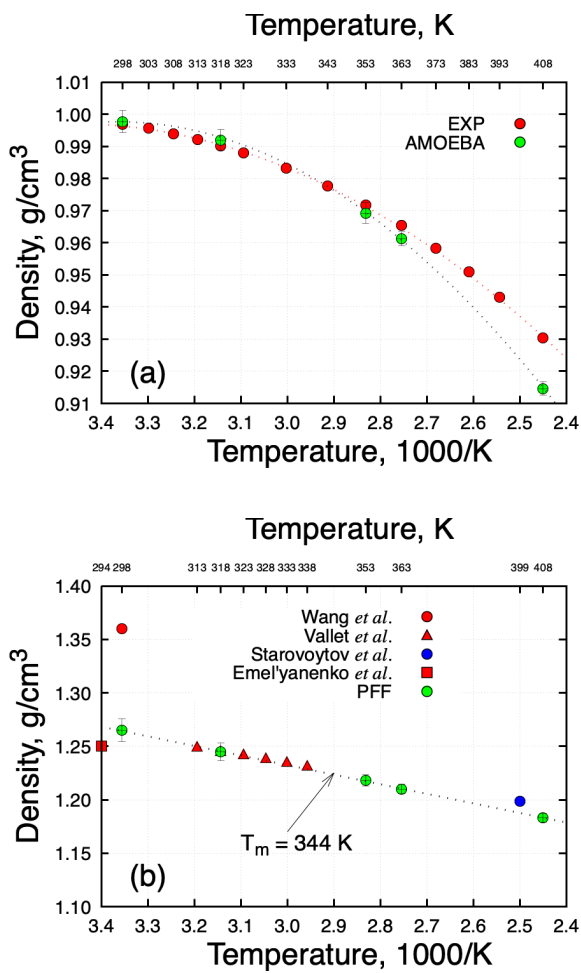

**Figure 4.** Liquid densities of water (a) and 1,3-dimethylimidazolium nitrate (b) as a function of temperature. Experimental data (red circles) was taken from available corresponding refs.[10,16,58,61,62] Experimental points in (b) correspond to 1,3-dimethylimidazolium nitrate [DMIM][NO$_3$] (red circle), 1-ethyl-3-methyl imidazolium nitrate [EMIM][NO$_3$] (red triangles) and 1-methylimidazolium nitrate [HMIM][NO$_3$] (red square). The black dashed arrows represent trend lines: second order polynomial (a) and linear (b).

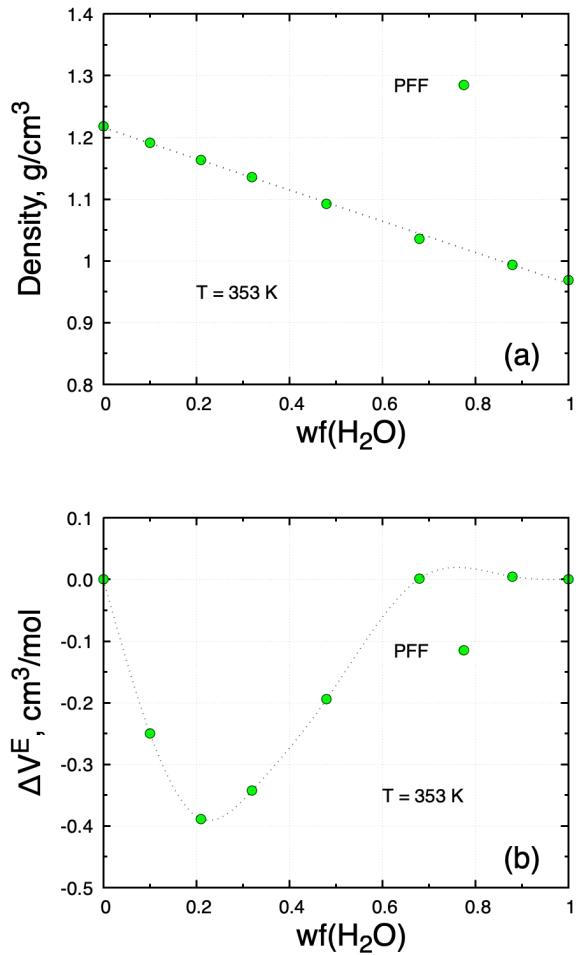

**Figure 5.** Liquid density ρ (a) and excess molar volume $\Delta V^E$ (b) for 1,3-dimethylimidazolium nitrate - water mixtures are shown as a function of the water weight fraction (wf) using a revised polarizable force field (black circles). The black dashed arrows represent trend lines: linear (a) and second order polynomial (b).

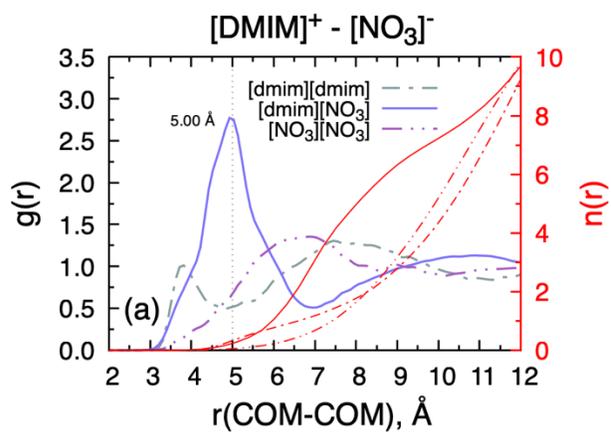
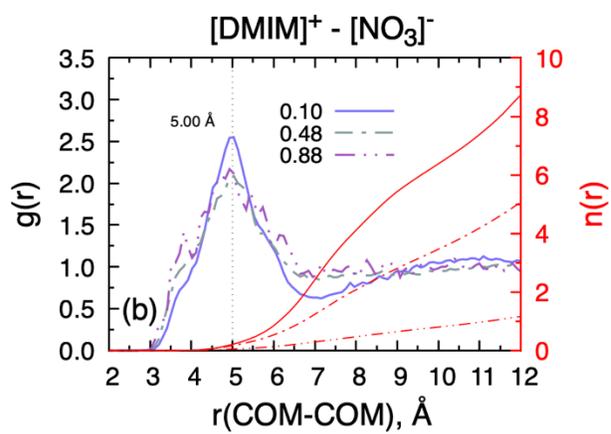

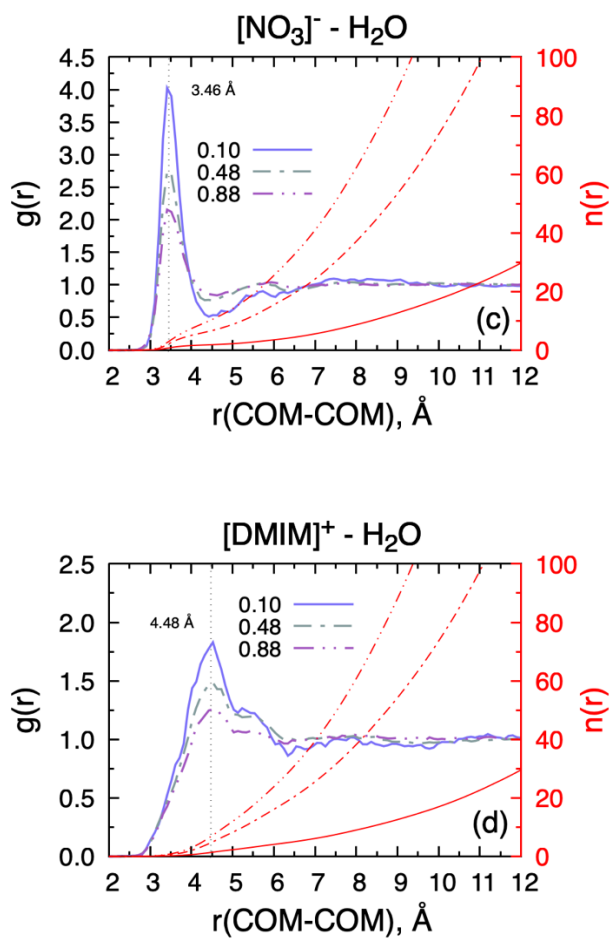

**Figure 6.** Center of mass radial distributions is given for the pure 1,3-dimethylimidazolium nitrate (a) and 1,3-dimethylimidazolium nitrate - water mixtures as a function of water weight fraction (b-d). There are three water weight fractions 0.10, 0.48, and 0.88. Radial distributions are plotted for [DMIM][NO$_3$] pair in (b), [NO$_3$]$^-$ - H$_2$O in (c), and [DMIM]$^+$- H$_2$O in (d). Coordination numbers (red color lines) are also shown for each considering pair.

**TABLES**

**Table 1.** Inter-molecular distances and corresponding energies for the cation/anion and anion/water pairs.

| Level of theory | r(C2-N), Å | E, kcal/mol | r(O-N), Å | E, kcal/mol |
|---|---|---|---|---|
| | [DMIM]$^+$ - [NO$_3$]$^-$ | | [NO$_3$]$^-$ - H$_2$O | |
| HF | 3.43 / 3.46 | -89.77 / -87.78 | 3.36 / 3.41 | -14.35 / -14.18 |
| MP2 | 3.35 / 3.40 | -92.25 / -89.17 | 3.28 / 3.32 | -15.56 / -15.50 |
| B3LYP | 3.38 / 3.42 | -92.81 / -89.13 | 3.28 / 3.32 | -16.36 / -16.18 |
| ωB97X-D | 3.36 / 3.40 | -95.43 / -92.18 | 3.27 / 3.30 | -17.33 / -17.34 |
| B3LYPD3 | 3.16$^*$ / 3.41 | -97.08 / -92.26 | 3.27 / 3.31 | -18.02 / -17.83 |

Inter-molecular distances and corresponding energies are given for 6-311G(d,p) / 6-311++G(d,p) basis sets.
Corresponding energies are calculated with BSSE correction.
$^*$ Dimer geometry is not planar

**Table 2.** Nonbonded parameters and atomic polarizabilities for the nitrate anion $[NO_3]^-$.

| Atom type | Atom class | $R_{min}$, Å | $\varepsilon$, kcal/mol | $\alpha$, Å$^3$ |
|---|---|---|---|---|
| N | 94 | 3.710 | 0.105 | 1.073 |
|   |    | (3.550) | (0.265) | (0.823) |
| O | 95 | 3.300 | 0.112 | 0.837 |
|   |    | (3.260) | (0.210) | (0.405) |

( ) data in parenthesis were taken from Starovoytov *et al.*, JPCB, 2021, 125, 40, 11242-11255

**Table 3.** Liquid volumes (V) for the 1,3-dimethylimidazolium nitrate pair using GDMA G1 multipoles

| T, K | $\langle V \rangle$, Å$^3$ |
|---|---|
| 353 | 216.18 |
| 363 | 217.66 |
| 408 | 223.69 |

**Supporting Information**

The following files are available free of charge at …

A revised dihedral potential, optimized geometries for [DMIM][NO$_3$], [DMIM]$^+$/H$_2$O, and [NO$_3$]$^-$/H$_2$O, total inter-molecular potentials, center-of-mass radial distributions for ionic liquid/water mixtures.

AUTHOR INFORMATION

**Corresponding Author**

**Oleg N. Starovoytov** – Department of Computer Science, Southern University and A&M College, Baton Rouge, Louisiana 70813, United States; https://orcid.org/0000-0002-5784-7312;

E-mail: oleg_starovoytov@subr.edu

**Author Contributions**

* E-mail: oleg_starovoytov@subr.edu

**Phone : +1-517-858-9394**

**Notes**

The author declares no competing financial interest.

ACKNOWLEDGMENT

The author expresses acknowledgements to the Extreme Science and Engineering Discovery Environment (TG-DMR140104) and the Louisiana Optical Network Initiative (LONI). Acknowledgements are expressed to Prof. Shizhong Y. for the collaboration and help for completing this project.

ABBREVIATIONS

Quantum mechanics (QM), molecular mechanics (MM), molecular dynamics (MD), the Hartree-Fock method (HF), the Møller-Plesset second order perturbation theory (MP2), the coupled cluster complete-basis level CCSD(T), the Kohn-Sham formalism (KS), the general gradient approximation (GGA), basis set superposition error (BSSE), the counterpoise correction approach (CP), a polarizable force field (PFF), room temperature ionic liquids (RTILs), a hydrogen bond (HBD), a periodic simple cubic lattice (SC), and nanosecond (ns).